\documentclass[aps,showpacs,amsmath,amssymb,floatfix]{revtex4}

\usepackage{pslatex}
\usepackage[dvips]{graphicx} 
\usepackage{amssymb}
\usepackage{url}
\begin{document}

\title{From heavy ions to exotic atoms} 
\author{ P. Indelicato}
\email{paul@spectro.jussieu.fr}
\author{ M. Trassinelli}
\affiliation{
Laboratoire Kastler Brossel, \'Ecole Normale Sup\'erieure et Universit\'e P. et M. Curie, Case 74, 4 place Jussieu, F-75252, Cedex 05, France}

 \begin{abstract}We review a number of experiments and theoretical calculations on heavy ions and exotic atoms,
which aim at providing informations on fundamental interactions. Among those are propositions of experiments for
parity violation measurements in heavy ions and high-precision mesurements of  He-like transition energies in highly charged ions.
We also describe recent experiments on pionic atoms, that make use of highly-charged ion transitions to obtain
accurate measurements of strong interaction shift and width.
\end{abstract}

 \maketitle

\setcounter{page}{1}

\section{Introduction}\label{intro}

In the last 30 years, accelerators and heavy ion sources around the world have been used
to produce highly charged heavy ions to provide accurate tests of quantum electrodynamics
in strong Coulomb fields. Indeed, the work of Lamb and Retherford \cite{lar1950} 
and of Kush and Foley \cite{kaf1948} have been at the origin
of quantum Field Theories in the first place.  But atomic physics experiments have proven valuable also as a tool to study fundamental 
physics way beyond the realm of Quantum Electrodynamics. In particular the study of parity violation in atoms
\cite{bab1974} provided constraints on the electro-weak model, which for a while were completing the
LEP results,  with a comparable accuracy \cite{wbc1997}. Accurate measurements of the electron electric dipole moment in atoms are also providing interesting 
limits to time-reversal violation (see, e.g., \cite{gaf2004} for a recent review). 

Exotic atoms and ions are formed when a heavy particle is captured on an atomic nucleus. In the process, many or all electrons of the atom are ejected and one 
obtains a highly charged heavy ion, with properties that depend on the captured particle.
Measurements in exotic atoms or ions have lead to accurate long-lived particle mass measurements, and to strong-interaction studies (using 
pionic, kaonic and antiprotonic atoms) at very low energy. Muonic atoms have also been used for precise studies of nuclear structure
(nuclear charge distribution radius and shapes). Finally the availability of antiprotonic helium atoms \cite{ymh2002} and of antihydrogen \cite{aab2002,gbo2002} is leading
to new tests of the CPT theorem.

In this paper we review some theoretical and experimental work, that exemplifies this aspect of atomic physics, restricting ourselves to problems concerning heavy ions and exotic
atoms.
In Sec. \ref{sec:pnc}, we describe proposals that concern parity violation in heavy ions.  In Sec. \ref{sec:pih}, we will show how the physics of exotic atoms and of highly charged heavy ions have come together to provide
at the same times accurate spectrometer calibrations that were mandatory to obtain a significant result on the strong interaction broadening in pionic hydrogen, while showing
prospects for very accurate measurements with highly-charged heavy ions.

\section{Parity violation in heavy ions}\label{sec:pnc}  
Parity violation in atoms results from the exchange of $Z_0$ bosons (the so-called neutral currents) between atomic electrons and nucleons in the nucleus.
From that assumption, one can derive the Hamiltonian  
\begin{equation}
\label{eq:hpnc}
H_{\rm pv} = \frac{G_F}{2 \sqrt{2}}
             (1 - 4 \sin^2 \vartheta_{\rm W} - \frac{N}{Z})
                              \rho  \gamma_5 
\,.
\end{equation}
where $G_F$ denotes Fermi's constant, $\vartheta_W$ the Weinberg angle, $N$
the neutron number, $Z$ the proton number, and $\rho$ the nuclear density
normalized to $Z$ and $\gamma_5=i\gamma_0\gamma_1\gamma_2\gamma_3$. The $\gamma_{\mu}$ are Dirac matrices. This formula also demonstrates why highly charged heavy 
ions with few electrons are proper candidates for investigating
parity non--conservation effects: The wave function admixture coefficient 
$\eta_{\rm pv}$ between level $i$ and $f$, which is given by
\begin{equation}
\label{eq:pncamp}
\eta_{\rm pv} = \frac{\langle i | \frac{G_F}{2 \sqrt{2}}
             ( 1- 4 \sin^2 \vartheta_{\rm W} - \frac{N}{Z} )
                              \rho \gamma_5   | f \rangle}{E_i - E_f}
= \frac{\langle i |H_{\rm PV} |f \rangle}{E_i - E_f}\,,
\end{equation}
is very large (typically orders of magnitude larger than for
the outer shell in neutral atoms) due to the big overlap
between the nucleus and the electron states. In order to get even larger enhancement, one has
to find quasi-degenerate states, so that $\left| E_i - E_f\right|$ is very small. This is a rather general scheme. For example, the
 coupling by the hyperfine Hamiltonian of a long-lived and of a short-lived 
state can lead to very large effects on the short-lived state lifetime, a phenomenon known as hyperfine quenching. The use of the quasi-degeneracy between the $1s2p\, ^3P_0$ and $1s2p\, ^3P_1$ fine -structure states
 in heliumlike ions,  as proposed in 1989 \cite{ipm1989} has lead
to a very fruitful study of the fine structure splitting over a large range of atomic numbers\cite{gmm1974,msid1989,dllb1991,ibbc1992,bbcd1993,tmbb2004}.

\begin{figure}[htb]
\centering
\includegraphics[height=6cm]{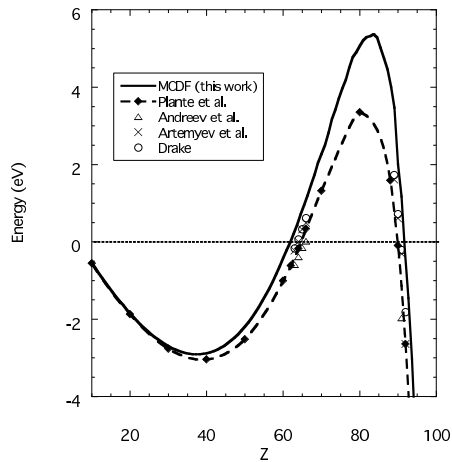}
\caption[]{comparison of different calculations of the $1s2p\, ^3P_0$ and $1s2s\, ^1S_0$ energy splitting. MCDF: \cite{ind1988,ind1995,msgi1996} and this work;
 Plante \textit{et al}: \cite{pjs1994}; Artemyev \textit{et al}: \cite{asyp2005}; Andreev  \textit{et al}: \cite{alps2003}; Drake: \cite{dra1988}}
\label{he-degen}
\end{figure}

 The search for a similar enhancement in the case of parity violation has lead to a proposal \cite{ssim1989} to use the accidental quasi-degeneracy between the 
$1s2p\, ^3P_0$ and $1s2s\, ^1S_0$ opposite parity levels in Heliumlike ions for $Z\approx 92$.
In that case, $\left| E_i - E_f\right| \approx 1$~eV can be very small compared to the energy of the transitions to the ground state ($\approx 100$~keV) or to the $n=2$ shell binding energy ($\approx 25$~keV).
The idea was to compare the rate of the parity-violating $2E1$ transition between the $1s2p\, ^3P_0$ and $1s2s\, ^1S_0$, excited by a laser, to the $E1M1$ rate. At the time
the required laser intensity ($10^{21}$~W/cm$^2$), obtained with a theoretical $1s2p\, ^3P_0$ and $1s2s\, ^1S_0$ energy splitting of -0.9~eV, were out of reach.
 There is still a large uncertainty on the energy splitting. A summary of the different calculations is presented on Fig.~\ref{he-degen}. The value of $Z$ closest to the
crossing point can vary by one or two units, and the uncertainty,  due in particular to nuclear finite size and shape corrections are large. A direct measurement of this 
splitting for several elements would be very valuable as it would provide simultaneously information on QED corrections and nuclear structure.

More recently, a number of other proposals addressing other observables have been proposed, either on Be-like ions \cite{msi1998}, or making use again of the quasi-degeneracy in
heliumlike ions  mentioned above
but this time at the other crossing, around gadolinium and europium \cite{lnps2001}.

\section{The strong interaction shift and broadening in pionic hydrogen}
\label{sec:pih}
\subsection{The new pionic hydrogen experiment}
Exotic atoms are atoms that have captured a long-lived, heavy
particle. This particle can be a lepton, sensitive only to the
electromagnetic and weak interactions, like the electron or the muon,
or a meson like the pion, or a baryon like the antiproton. An other
kind of exotic atom is the one in which the \emph{nucleus} has been
replaced by a positron (positronium, an e$^+$e$^-$ bound system) or a positively
charged muon (muonium, a $\mu^+$e$^-$ bound system).

The capture of a negatively charged, heavy particle $X^-$ by an atom,
occurs at a principal quantum number $n\approx n_e
\sqrt{\frac{m_{X^{-}}}{m_e}}$ where $n_e$ is the principal quantum number of
the atom outer shell, and $m_e$, $m_{X^-}$ are respectively the
electron and particle mass.  This leads to $n=14$, 16 and 41 for the capture on hydrogen or helium of
muons, pions and antiprotons respectively. The capture process populates
$\ell$ sub-states more or less statistically. During the capture process
of an heavy, negatively charged particle, many or all of the electrons
of the initial atoms are ejected by Auger effect. As long as electrons
are present, Auger transition rates are very large and photon emission
is mostly suppressed except for the low lying states. For light elements,
 or very heavy particles like the
antiproton, the cascade can end up with an hydrogenlike ion, with only
the exotic particle bound to the nucleus \cite{sabb1994}.

The spectroscopy of exotic atoms has been used as a tool for the study
of particles and fundamental properties for a long time. Exotic atoms
are also interesting objects as they enable to probe aspects of atomic
structure that are quantitatively different from what can be studied
in electronic or ``normal'' atoms. For example, all captured particles
are much heavier than the electron, and thus closer to the nucleus,
leading to a domination of vacuum polarization effects over
self-energy contributions, in contrast to normal atoms.

Pions are mesons, i.e., particles made of a quark-antiquark
pair. They are sensitive to strong interaction. They decay into a muon and a muonic neutrino.  The lifetime of the charged pion is $2.6\times
10^{-8}$~s. The mass of
the pion is 273 times larger than the electron mass. Contrary to the
electron, it has a charge radius of $\approx 0.8$~fm and is a spin-0 boson.

Quantum Chromodynamics is the theory of quarks and gluons, that
describe the strong interaction  in the
Standard Model. It has been studied extensively at high-energy, in the
asymptotic freedom regime, in which perturbation theory in the
coupling constant can be used. This does not work in the low-energy limit. Weinberg proposed
Chiral Perturbation Theory (ChPT) \cite{wei1979} to deal with this
problem. More advanced calculations have been performed since then,
that require adequate testing (see, e.g., \cite{gilm2003}). The lifetime of the 
pionium (a bound pion-antipion system) has been studied at CERN \cite{aabb2004}, because it is the most elementary object for studies of ChPT. Yet 
the accuracy is not very good because of the difficulty of the experiment, and pionic
hydrogen is the best candidate for accurate test of ChPT. The shift
and width of np$\to$1s transition in pionic hydrogen due to strong
interaction can be connected respectively to the
$\pi^-$p$\to\pi^-$p and $\pi^-$p$\to\pi^0$n cross-sections,
which can be evaluated by ChPT, using a Deser-type
formula \cite{dgbt1954}.
 
Pionic hydrogen has thus been subjected to a 10 year experimental program \cite{sbgj2001}, completed a few years ago, that has yielded a 9~\% accuracy for the strong interaction 
broadening (of the order of 1~eV) and a 0.6~\% accuracy on the energy shift (of the order of 7~eV). Yet, after such an effort, the
knowledge of the width was not good enough for an accurate test of ChPT and it was worthwhile to consider new ways of doing the experiment \cite{naab2002,abbd2003,mabb2003,got2004}.

\begin{figure} [!hbp]
\centering
\includegraphics[clip=true,scale=0.5,angle=0]{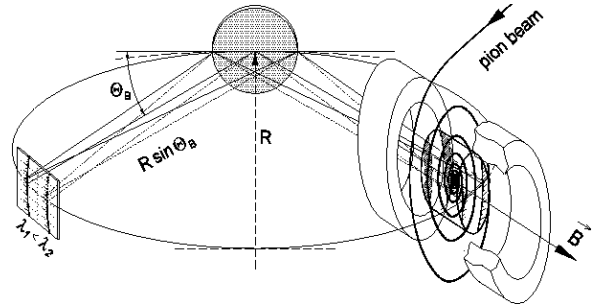}
\caption{Principle of X-ray spectrocopy of exotic atoms with the
cyclotron trap and a spherically curved crystal spectrometer. The
bidimensional X-ray detector is a 6-chips cooled CCD detector and is
located on the Rowland circle of radius $R/2$, where $R$ is the radius
of curvature of the crystal ($\approx 3$~m)
}\label{fig:trap}
\end{figure}

Since the mid-ninety a large collaboration has developed a technique to 
measure with great accuracy  X-ray spectra from light exotic atoms, formed in a low-density gas.
This techniques involves the cyclotron-trap, a small cyclotron that can decelerate  and focus a particle beam on 
a gas target located at its center, with the help of degrader foils. The second version of this device can efficiently decelerate
antiprotons and pions, and also secondary beams like muons, which originate in pions' decay. The techniques involves also
a high-resolution high-luminosity spherically-curved crystal spectrometer equipped with a 6-chip X-ray CCD camera, that enable a high level of
background rejection. A sketch of the experiment is presented on Fig.~\ref{fig:trap}. More details on the experimental set-up can be found in
\cite{gaab1999,naab2002,got2004}.
After a successful study of antiprotonic hydrogen and deuterium at LEAR (CERN) \cite{gaab1999}, and a measurement of the pion mass \cite{lbgg1998}, a
first attempt at studying pionic deuterium
provided in a very short time a sizable improvement over
previous experiments \cite{hksb1998}. It was then decided that such an
apparatus could lead to improvements in pionic hydrogen measurement of a factor 3
in the accuracy of the shift and of one order of magnitude in the
accuracy of the width. 

\subsection{He-like heavy ions and pionic hydrogen}
\label{sec:ions}

The main difficulty in the experiment is to separate the strong
interaction broadening of the pionic hydrogen lines from other
contributions, namely the instrumental response function, Doppler
broadening due to non-radiative de-excitation of pionic hydrogen
atoms by collisions with the H$_2$ molecules of the gas target and
from possible transitions in exotic hydrogen molecules.

The Doppler broadening and the role of exotic molecules can be estimated by systematic studies
 as a function of target density and of the initial level of the transition.
We have thus measured the width of the transitions $np\to1s$, with $n=2$, 3 and 4.   A detailed theoretical study
of the cascade processes has also been performed \cite{jen2004}.This calculations  provides the kinetic energy distribution
of  pionic hydrogen atoms in the different levels, thus enabling to correct the width. With that correction we obtain similar width from transitions
originating from all three $np$ levels.
We also measured similar transitions in muonic hydrogen, in a case where there is no strong interaction broadening,
to check our ability to analyze properly cascade and Doppler broadening.

The determination of the instrumental response, which depending of the crystal varies from
$\approx 0.28$~meV to $\approx 0.4$~eV, was difficult. X-ray emitted by neutral atoms in solid or gas target, ionized in inner shells
cannot be used because of their natural width ($\approx 1$~eV).  Exotic atoms do not
provide as good a response function calibration as most lines, coming
from molecules like N$_2$, are broadened by Doppler effect due to the Coulomb
explosion during the atom formation process \cite{sabg2000}. Finally
the rate is much lower and the statistic usually  not sufficient.

The instrumental response has then been studied using a transition in helium-like
ions \cite{abgg2005}, emitted by the plasma of a high-performance Electron-Cyclotron Resonance
Ion Trap (ECRIT) build at PSI \cite{bsh2000}, using the cyclotron trap and a permanent magnet hexapoles.

\begin{figure} [!hbp]
\centering
\includegraphics[clip=true,scale=0.5,angle=0]{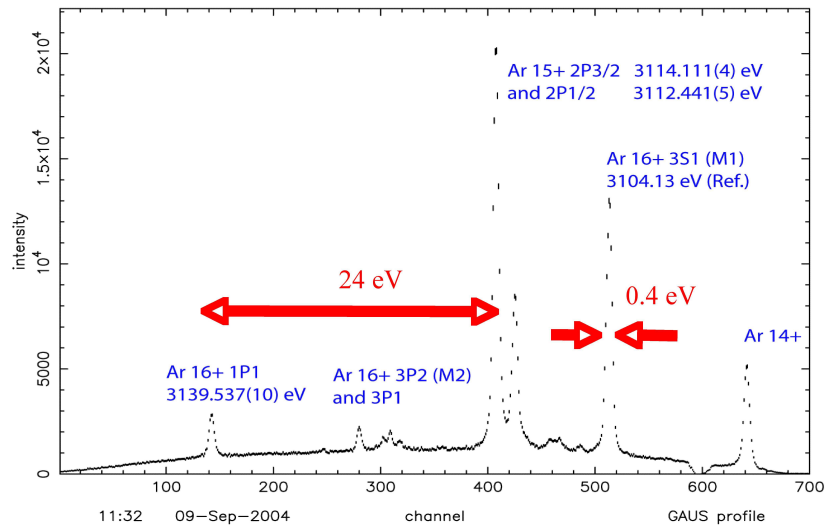}
\caption{High-statistic X-ray spectrum of helium-like argon (center
line) from the PSI Electron-Cyclotron Resonance Ion Trap, acquired using the instrument in Fig.~\ref{fig:trap}.  The width of the M1 
3.1~keV line is 0.4~eV. It is used as an energy reference.
}\label{fig:ar16}
\end{figure}

 With this ECRIT, we have obtained high-intensity spectra of highly charged sulfur, chlorine and argon. 
For these elements the relativistic M1 transition $1s2s \,^3S_1\to 1s^2 \,^1S_0$ is very bright and very narrow, due to the very low kinetic 
energy of the ions in the source (Doppler width $\leq 40$~meV, natural width is negligible). The M1 transitions in each
of these elements are located nearby the $n=2$, 3 and 4 transitions in pionic hydrogen.
They thus allow
for systematic study of instrumental response. An
example of the relativistic M1 transition spectrum in He-like argon is
presented on Fig.~\ref{fig:ar16}.

\subsection{High-precision measurements in highly charged ions}
Apart from the characterization of the spectrometer, one of the most important results of this ongoing project is in a series
of  high-precision measurement of the 
X-ray spectra of argon, chlorine and sulfur. With 1--2 hours  maximum
acquisition time, we obtained high-statistics spectra of {He-}, Li- and Be-like ionic states
of these elements. This new approach may lead to a better understanding of a fundamental problem that arose recently:
after several decades of high-precision work on He and helium-like ions, there are still difficulties in comparing theory and experiment.
In He there is a strong disagreement (six times the combined error bars) between the calculations \cite{dra1996,pas2000,pas2002,pas2003} and recent very accurate measurements
\cite{sgh2000,glh2001,pgdh2004}. In He and light elements, the theory makes use of an expansion of the Bethe and Salpeter equations in operators of
 successive orders in $\alpha$, the fine structure constant, and $Z\alpha$. For heavier elements one uses all-orders methods, in which series of diagrams are summed up,
 and  QED calculations to all orders in $Z\alpha$ are available. 
 One way  to understand this very important discrepancy, is then to study more accurately heavier elements.

A single crystal spectrometer like ours can only measure energy differences between atomic transitions. A reference is thus needed.
Due to the lack of high quality reference lines in neutral atom X-ray spectra (see, e.g., \cite{agis2003}) we used as a reference He-like 
$1s2s\;^{3}S_1 \to 1s^2\;^{1}S_0\; M1$ transition. All the transition energy provided in the present work are relative to the M1 theoretical transition energy.
The peaks in the spectra were fitted with a simulated spectrometer response function which was 
convoluted to a Gaussian.
The response function was obtained through a Montecarlo X-ray tracking simulation based on the
theoretical reflection function of the crystal obtained with the XOP code~\cite{sad1998}.
The reliability of the simulation had previously been tested during the crystal response 
function study~\cite{abgg2005}.
The results obtained in highly-charged argon and sulfur spectroscopy have an
unprecedented precision of the order of 10~meV and they agree with the previous 
experimental values and theoretical predictions (see Tables~\ref{argon_en},~\ref{sulfur_en}).

\begin{table}[h!]
\begin{tabular}{lll}
\hline
Transition:& \multicolumn{1}{r}{$1s2p\;^{1}P_1 \to 1s^2\;^{1}S_0$}
  & \multicolumn{1}{r}{$1s2p\;^{1}P_1 \to 1s2s\;^{3}S_1$}\\
\hline
  {Costa~\cite{cmps2001} (th.)}              & 3139.57      & 16.05\\
  {Plante~\cite{pjs1994} (th.)}            & 3139.6236    & 16.0484\\
  {Lindgren~\cite{lasm2001} (th.)}       &              & 16.048\\
  {Artemyev~\cite{asyp2005} (th.)} & 3139.5821(5) & 16.0477(2) \\
  {Deslattes~\cite{dbf1984} (exp.)} & 3139.553(36) & 16.031(72)\\
  {This Work}               & 3139.537(10) & 16.040(17)\\
  (prelminary results)\\

\hline
\end{tabular}
\caption{Example of experimental determination of He-like argon  transition energies in eV\cite{tbbc2005}}
\label{argon_en}
\end{table}

\begin{table}[h!]
\begin{tabular}{lll}
\hline
Transition: & \multicolumn{1}{r}{$1s2p\;^{1}P_1 \to  1s^2\;^{1}S_0$}
  & \multicolumn{1}{r}{$1s2p\;^{1}P_1 \to  1s2s\;^{3}S_1$}\\
\hline
 P.~Indelicato~\cite{ind1988,ind1995,PrivatePI96} (th.)                       & 2460.6169  & 13.4875\\
 Plante~\cite{pjs1994} (th.)                & 2460.6707  & 13.4857\\
 Artemyev~\cite{asyp2005} (th.) & 2460.6292(4) & 13.4853(2) \\
 Schleinkofer~\cite{sbbt1982} (exp.)& 2460.67(9) & 13.62(20)\\
 This Work                    & 2460.608(9)& 13.483(16)\\
 (prelminary results)\\

\hline
\end{tabular}
\caption{Example of experimental determination of He-like sulfur  transition energies in eV\cite{tbbc2005}}
\label{sulfur_en}
\end{table}

\section{Conclusions and perspectives}\label{concl}

In this paper, We have shown that the physics of highly charged ions can be used in several different circumstances to test fundamental
theories. We have first reviewed prospects of doing parity violation experiments with heliumlike ions. It is possible that with the FAIR project at GSI (\url{https://www.gsi.de/fair/index_e.html}), such experiments
become possible, but a large effort must be done to understand better the structure of two-electron ions.
 We have described a new set-up, which reuses instruments developed for exotic atoms, to build an ECRIT that generate highly-charged ions.
The high-performance X-ray spectrometer allows fro very accurate measurements of transition energy in two-electron ions, that could help solve the
problem in the fine structure of He. Thanks the high-intensity of the helium-like X-rays emitted by the ECRIT or by more conventional ECRIS (Electron-Cyclotron Ion Source), it is now 
foreseen that absolute energy measurements of at least M1 transitions are possible, using metrology-grade two-crystal instruments. Such an instrument, dedicated
to such experiments with highly charged ions is now being completed in our laboratory (\url{http://dirac.spectro.jussieu.fr/x-ray-metro.html}).

 On the exotic atom side, this set-up and the use of helium-like ions as high-quality X-ray standards, are leading
to highly improved measurements of the strong width and shift in pionic hydrogen and deuterium, as very precise tests of ChPT as a model of strong interaction
at low energy.

\section*{Acknowledgement(s)}
Laboratoire Kastler Brossel is Unit{\'e} Mixte de Recherche du
CNRS n$^{\circ}$ 8552. The experiments on pionic hydrogen are presented on behalf of
the pionic hydrogen collaboration (\url{http://pih.web.psi.ch}).

\bibliography{refs}
\vfill\eject
\end{document}